# Computational Investigation of Microstrip Antennas in Plasma Environment

Hardik Vyas, Bhaskar Chaudhury, Sanjeev Gupta
DA-IICT, Gandhinagar, 382007, Gujarat, India

*Abstract-* **Microstrip antennas are extensively used in spacecraft systems and other applications where they encounter a plasma environment. A detailed computational investigation of change in antenna radiation properties in the presence of plasma has been presented in this paper. The study shows antenna properties such as the resonant frequency, return loss, radiation properties and the different characteristics of the antenna changes when it is surrounded by plasma. Particular focus of the work is to understand the causes behind these changes by correlating the complex propagation constant in the plasma medium, field distribution on the patch and effective dielectric of the antenna substrate with antenna parameter variations. The study also provides important insights to explore the possibilities of designing tunable microstrip antenna where the substrate can be replaced with plasma and important antenna characteristics can be controlled by varying the plasma density.**

## I. INTRODUCTION

Microstrip patch antennas used in aerospace vehicles such as satellites, missiles etc. encounter plasma in space and during re-entry into the atmosphere. In the atmosphere, inhomogeneous plasma is formed around the vehicle due to friction between air and space vehicle [1]. Similarly, wireless sensors with patch antennas used for status monitoring of high voltage power lines (in power grids) encounters air plasma. The plasma so formed in both the cases acts as a dispersive and lossy dielectric which attenuates and changes the properties of electromagnetic (EM) waves originating from the antenna. Important antenna parameters such as resonant frequency, input impedance, return loss (RL), efficiency, bandwidth etc. are altered due to plasma thereby degrading its performance [1]. Therefore it is important to understand in details the effect of various plasma parameters on patch antenna especially due to its narrow bandwidth.

Past research in this area has been mostly focused on theoretical investigations and is based on the understanding of EM wave plasma interaction. Most of the theoretical and computational work is established on the assumption of collisionless homogeneous plasma layer on microstrip antenna [1],[2],[3]. However, in real scenario plasma is inhomogeneous and collisional which can substantially affect the antenna properties. To the best of our knowledge, the past computational studies are mostly observational in nature and limited plasma parameters have been taken into account. The objective of our work is to understand in details the effect of all important plasma parameters on radiation characteristics taking into account the plasma inhomogeneity and collisional behavior of plasma at high pressures.

As a first step, this paper presents a parametric computational study of the effect of a homogeneous collisional plasma layer on a patch antenna and the causes behind the changes in antenna parameters such as efficiency, resonant frequency, bandwidth, return loss (RL) etc. Two most important plasma parameters which are taken into consideration for this computational investigation are: change in electron density ($n_e$) and plasma thickness ($d$). Subsequently the simulations are performed to understand the effect of inhomogeneous plasma layer on antenna parameters.

## II. PLASMA EM WAVE INTERACTION

To understand the problem in details, it is important to revisit the theory of interaction of EM wave with plasma [4],[5]. Plasma, considered as the fourth state of matter, is a collection of ions, electrons and neutral gas particles. In plasmas, electrons being the most mobile species undergo oscillations about their mean position with a frequency called as plasma frequency $\omega_p$. Electrons also collide with other plasma particles particularly with neutral gas particles with the collision rate given by $v_c$ which varies with pressure. For cold collisional plasma, complex relative permittivity is given as [1],[4] :

$$\epsilon_p(\omega) = 1 + \frac{\omega_p^2}{\omega(jv_c - \omega)} \quad , \text{where } \omega_p = \sqrt{\frac{n_e e^2}{\epsilon_o m_e}} \quad (1)$$

$\epsilon_o$ is free space permittivity, $\omega$ is wave frequency, $n_e$ is the plasma density, $e$ and $m_e$ are electron charge and electron mass respectively.

The propagation constant for an EM wave in a collisional plasma is given by [4]

$$k(\omega) = \frac{\omega}{c}\sqrt{1 - \frac{\omega_p^2}{\omega(\omega - jv_c)}} \quad (2)$$

where c is the speed of light in air. The normalized real ($k_r$) and imaginary ($k_i$) components of $k$ with respect to free space wave number $k_o$ can be written in terms of $\omega_p/\omega$ and $v_c/\omega$. The real part $k_r$ is responsible for the phase shift whereas $k_i$ takes into account the attenuation of EM wave.

The 2D surface plots of $k_r/k_o$ and $k_i/k_o$ w.r.t $\omega_p/\omega$ and $v_c/\omega$ shown in Fig.1 gives a detailed understanding of the important parameter space. From the plot we can see that collision frequency and plasma frequency have significant effect on real part of "k" when $0.45 \omega \leq \omega_p \leq \omega$ and $0.1\omega \leq v_c \leq 2\omega$. In this region, $k_r/k_o$ decreases from 0.9 to 0.5 indicating the possibility of considerable change in phase. Change in $k_r$ may affect the antenna radiation pattern in case of an inhomogeneous plasma layer. Imaginary component $k_i$ represents the attenuation of EM wave due to collisional losses in the plasma medium. For $\omega_p \geq 0.4\,\omega$, $k_i/k_o$ is maximum (0.3) when collision rate and plasma frequency are comparable and it decreases with further increase or decrease in $v_c/\omega$. So plasma behaves as a lossy dielectric for the region discussed when collision frequency is comparable with the wave frequency. Increasing $\omega_p$ makes plasma more conductive leading to high attenuation and reflection of EM waves from plasma takes place at plasma frequencies greater than the wave frequency.

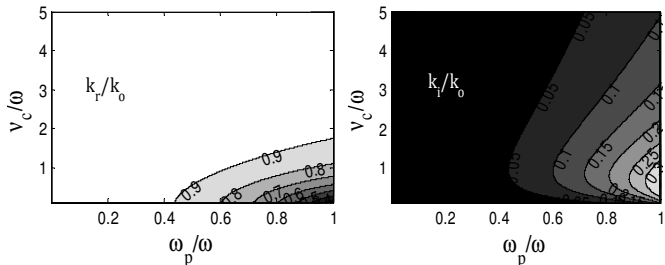

Fig.1. 2D surface plots of $k_r/k_o$ and $k_i/k_o$ vs $\omega_p/\omega$, $v_c/\omega$

### III. COMPUTATIONAL ANALYSIS AND RESULTS

For the simulations presented in this paper, we have considered a rectangular microstrip antenna with a line feed as described in ref. [6]. ANSYS HFSS is used based on Finite Element Method, for all our simulations. The antenna is designed for frequency f=7.5 GHz ($\omega$=47.1x$10^9$ rad/sec) with patch length $L$=12.45 mm ($\lambda/2$), width $W$ =16mm and RT Duroid substrate (dielectric constant $\epsilon_r$=2.2, $\tan\delta$=0.0009) with height h=0.795mm.

For parametric studies we considered two main cases, firstly the top surface of the antenna is covered by – a homogeneous plasma layer of thickness $d$ (case a, b) and secondly the top surface is covered by inhomogeneous plasma layer with thickness $d = 5h$ (case c). The inhomogeneous layer (case c) follows Epstein electron number density profile [4] with peak density of $4 \times 10^{17} m^{-3}$. The inhomogeneity was taken into account using 10 layers with respective electron densities from $2.06 \times 10^{17} m^{-3}$ to $1.7 \times 10^{17} m^{-3}$ with a difference of $0.04 \times 10^{17} m^{-3}$ between consecutive layers. The variations in $n_e$ has been taken into account by corresponding changes in permittivity, loss tangent and conductivity.

TABLE I: EFFECT OF PLASMA PARAMETERS ON ANTENNA PROPERTIES. COLLISION FREQ. $v_c$=12.87 $GHz$ and (a) $\omega_p$ =28.2x$10^9$rad/sec, (b) $\omega_p$ =35.6x$10^9$rad/sec

| Medium above the patch antenna | | RL (dB) | Res. Freq. (GHz) | BW % | Efficiency % |
|---|---|---|---|---|---|
| **Air** | | -26.1 | 7.5 | 2.80 | 96.2 |
| (a) **Plasma** ($n_e$) $2.5 \times 10^{17} m^{-3}$ | $d = h$ | -20.1 | 7.58 | 3.16 | 71.2 |
| | $d = 5h$ | -17.0 | 7.61 | 3.40 | 62.0 |
| (b) **Plasma** ($n_e$) $n_e 4 \times 10^{17} m^{-3}$ | $d = h$ | -15.0 | 7.62 | 3.60 | 57.4 |
| | $d = 5h$ | -13.2 | 7.67 | 3.25 | 46.8 |
| (c) **Inhomogeneous plasma** | $d = 5h$ | -18.7 | 7.60 | 3.15 | 66.3 |

Some of the observations from our study have been summarized in Table I and Fig. 2. A general trend was observed that the resonant frequency increases with increase in $n_e$ or increase in $d$. Increase in $n_e$ or increase in $d$ increases the amount of fringing and the antenna must resonate at a frequency greater than 7.5 GHz to get the required phase shift $k_r$ as shown in Fig.1. Efficiency decreases considerably with increase in $n_e$ or with an increase in $d$ due to increasing attenuation of EM wave in both the cases (Fig.1). It was found that in comparison to air, bandwidth increases in the presence of plasma. In the presence of plasma RL depends on plasma parameters because of possible impedance mismatch. Thus the antenna parameters vary in accordance with the EM wave parameters for both homogeneous as well as inhomogeneous plasma environment.

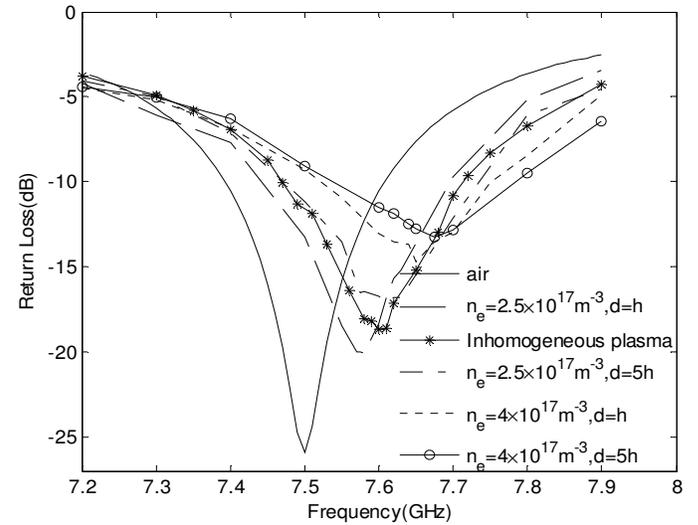

Fig.2. Return Loss vs Frequency for microstrip antenna covered with air / plasma with different parameters.

### CONCLUSIONS

Computational analysis of a rectangular microstrip antenna in a homogeneous, collisional plasma environment followed by an inhomogeneous environment has been carried out. The study predicts the general trends of change in antenna properties such as the resonant frequency, return loss, bandwidth, efficiency etc., with variations of important plasma parameters such as plasma density and plasma thickness. We have presented a qualitative as well as quantitative analysis of the changes of the antenna properties and the reasons behind the important changes.

### REFERENCES

[1] Z.H.Qian and R. S. Chen, "FDTD analysis of microstrip patch antenna covered by plasma sheath," *Progress in Electromagnetics Research*, vol. 52, pp.173–183, 2005.
[2] I.J.Bahl, , P. Bhartia, and S.S Stuchly, "Design of Microstrip Antennas Covered with a Dielectric Layer", *IEEE Trans. Antennas Propagat.* ,vol. AP-30, no. 2 , pp.314-317, March 1982.
[3] A.A.Sawalha. "Effect of ionized plasma medium on radiation properties of rectangular microstrip antenna printed on ferrite substrate". *International Journal of Physics and Astronomy*, vol. 2, No. 2,pp.65-77, June 2014.
[4] B. Chaudhury and S. Chaturvedi, "Comparison of Wave Propagation Studies in plasma using 3-D FDTD and Ray Tracing method". *Physics of Plasmas,* 13, 123302, (2006).
[5] T.H. Stix, *Waves in plasmas*. American Inst. of Physics, New York, 1992.
[6] D. M. Sheen et al. "Application of the Three dimensional FDTD method to the analysis of planar microstrip circuits", *IEEE Trans. on Microwave Theory Tech*, vol. 38, no. 7, pp. 849-857, July 1990.